\documentclass[twocolumn,aps,nofootinbib]{revtex4} 
\usepackage{graphicx, epsfig}

\newcommand{\be}{\begin{equation}}
\newcommand{\ee}{\end{equation}}
\newcommand{\bea}{\begin{eqnarray}}
\newcommand{\eea}{\end{eqnarray}}

\newcommand{\ApJ}{{\it Astrophys. J.\,}}

\newcommand{\etal}{{\it et al.}}

\def\fun#1#2{\lower3.6pt\vbox{\baselineskip0pt\lineskip.9pt
        \ialign{$\mathsurround=0pt#1\hfill##\hfil$\crcr#2\crcr\sim\crcr}}}

\renewcommand\({\left(}
\renewcommand\){\right)}
\renewcommand\[{\left[}
\renewcommand\]{\right]}

\newcommand\eq[1]{Eq.~(\ref{#1})}

\newcommand\GeV{\,\mbox{GeV}}

\newcommand\Mpc{\,\mbox{Mpc}}

\newcommand\mpl{M_{\rm P}}

\newcommand\lsim{\mathrel{\rlap{\lower4pt\hbox{\hskip1pt$\sim$}}
    \raise1pt\hbox{$<$}}}
\newcommand\gsim{\mathrel{\rlap{\lower4pt\hbox{\hskip1pt$\sim$}}
    \raise1pt\hbox{$>$}}}

\newcommand\diff{\mbox d}

\def\dslash{\not{\hbox{\kern-2pt $\partial$}}}
\def\Dslash{\not{\hbox{\kern-4pt $D$}}}
\def\Oslash{\not{\hbox{\kern-4pt $O$}}}
\def\Qslash{\not{\hbox{\kern-4pt $Q$}}}
\def\pslash{\not{\hbox{\kern-2.3pt $p$}}}
\def\kslash{\not{\hbox{\kern-2.3pt $k$}}}
\def\qslash{\not{\hbox{\kern-2.3pt $q$}}}

 \newtoks\slashfraction
 \slashfraction={.13}
 \def\slash#1{\setbox0\hbox{$ #1 $}
 \setbox0\hbox to \the\slashfraction\wd0{\hss \box0}/\box0 }
 
\def\ee{\end{equation}}
\def\be{\begin{equation}}

\def\calp{{\cal P}}
\def\calr{{\cal R}}
\def\calpr{{\calp_\calr}}

\newcommand\sub[1]{_{\rm #1}}

\newcommand\Tr{{\rm Tr}\,}

\begin{document}
\preprint{CERN-PH-TH/2004-148}
\setlength{\unitlength}{1mm}
\title{The running-mass inflation model and WMAP}
\author{Laura Covi$^1$, David H.~Lyth$^2$, Alessandro Melchiorri$^3$
and Carolina J.~Odman$^3$}
\affiliation{ 
$^1$ Theory Division, Department of Physics, CERN, CH 1211 Geneva,
Switzerland.\\
 $^2$ Physics Department, Lancaster University,
Lancaster LA1 4YB, United Kingdom.\\
$^3$ Physics Department, University of Rome ``La Sapienza'',
Ple Aldo Moro 2, 00185 Rome, Italy}
\date{\today}%

\begin{abstract}
We consider the observational constraints on the running-mass 
inflationary model, and in particular on the scale-dependence 
of the spectral index, from the new Cosmic Microwave Background 
(CMB) anisotropy measurements performed by WMAP and from new 
clustering data from the SLOAN survey. We find that the data 
strongly constraints a significant positive scale-dependence of $n$, 
and we translate the analysis into bounds on the physical parameters 
of the inflaton potential. 
Looking deeper into specific types of interaction
(gauge and Yukawa) we find that the parameter space is significantly 
constrained by the new data, but that the running mass model remains 
viable.
\end{abstract}
\bigskip

\maketitle 

\section{Introduction}

The measurements of the Cosmic Microwave Background (CMB) anisotropies 
provided by the Wilkinson Microwave Anisotropy Probe (WMAP) 
mission \cite{Bennett03} have truly marked the beginning of the era of 
precision cosmology. In particular, the shape of the measured 
temperature and polarization angular power spectra are
in spectacular agreement  with the expectations of the standard 
model of structure formation, based on primordial adiabatic and nearly scale 
invariant perturbations. 
Assuming this model of structure formation, accurate albeit 
{\it indirect} constraints on several cosmological parameters have been 
reported \cite{spergel} in agreement with those previously indicated 
(see e.g. \cite{cjo}) but with much larger error bars. 

Moreover, new, complementary, results from 
the Sloan Digital Sky Survey (SDSS) on galaxy clustering 
(see e.g. \cite{Tg04}) and, more
recently, on Lyman-alpha Forest clouds (\cite{Se04}) are now
further constraining the scenario.

The question naturally arises if this new cutting-edge cosmological 
data can tell us something about inflation, assuming of course that 
the vacuum fluctuation of the inflaton generates the 
primordial perturbation. (If instead some `curvaton' field 
\cite{curvaton1,curvaton2} does the job, the data give information 
mostly about the properties of the curvaton, 
during some post-inflationary era.)
It is therefore not a surprise that several recent works have 
attempted to use this new cosmological data to
constraint and/or falsify inflationary physics.
In particular, constraints have been placed on 
the general form of the single field inflationary
potential, for instance in \cite{WMAP-peiris}, \cite{kkmr}, 
\cite{SL03} and \cite{Se04}.
A different approach is to ask about constraints on 
specific well-motivated models, constructed in accordance
with present ideas about what lies beyond the Standard Model of the
interactions. A comprehensive survey of such models has been provided
\cite{treview,book}, which in general predict an almost scale-invariant
spectral index $n<1$. Unfortunately, the 
data do not yet discriminate between these scale-invariant models 
\cite{Se04}.

In this paper, we study instead a specific inflationary scenario,
Stewart's running-mass model \cite{st97,st97bis,clr98,cl98,c98,rs}, 
a type of inflationary model which emerges naturally in the 
context of supersymmetric extensions of the Standard Model.
The model is of the single-field type (i.e.\ the slowly-rolling
inflaton field has only one component \cite{treview,book}), 
but nevertheless it has the striking signature of a very specific and
relatively strong scale dependence of the spectral index.
It is therefore a natural question to ask, if such a
dependence is compatible or even preferred by the
present data: in fact the WMAP collaboration claimed
to have a slight evidence for non-vanishing running
of the spectral index in their first year data. 
Even if their best fit value for $n'$ is too strong to be
accommodated in the usual paradigm of slow-roll models
and probably is generated mostly by the Lyman $\alpha$~\cite{Lya-run}
or even the low multipole data, we will attempt a 
conservative comparison and repeat our analysis 
in \cite{clm03}, on order to test the predictions of
the running-mass model.

Our paper is organized as follows: In section II we discuss the 
running mass model. In section III we present our data analysis 
method and results. Finally, in section IV, we discuss our conclusions.

\medskip
\section{The Running Mass Model}
\medskip

\subsection{The inflationary potential}

In common with all supersymmetric models, the running mass model
 \cite{st97,st97bis,clr98,cl98,c98,rs} chooses for the inflaton 
$\phi$  a flat direction, in order to suppress all the renormalizable
inflaton couplings. Also, $\phi$ is  many orders of magnitude less 
than  the reduced Planck scale $\mpl=2.4\times 10^{18} \GeV$ 
which ensures that the Planck-suppressed non-renormalizable terms 
are negligible.
Symmetries can moreover guarantee that odd powers of $\phi$ and 
in particular the possible lower order linear term~\cite{bcd} are absent.
The tree-level potential is therefore a constant $V_0$  plus a 
soft supersymmetry breaking mass term. 
Note then that in the running-mass model, inflation takes place 
in a region of field space in which  one (or more) of the fields 
are strongly displaced from the vacuum, typically by an amount 
$\sim \mpl$. The model is typically a realization of Linde's 
hybrid inflation~\cite{linde91}, where
 the displaced field is a `waterfall field', 
different from the slowly-rolling inflaton, and stabilized 
temporarily during inflation by an effective mass term.
At tree-level, the running-mass  model then reduces to the version 
of hybrid inflation proposed in~\cite{rsg}, distinguished from 
more general  models by the fact that the waterfall field is 
displaced from the vacuum by an  amount of order $\mpl$,
and it has also a mass not far above the generic
supergravity-mediated supersymmetry breaking value.

So in this setting, the inflaton potential is dominated simply 
by the soft supersymmetry-breaking mass term generated by $V_0$
and its radiative corrections. These are taken into account by 
using the Renormalization Group (RG) improved potential.
To construct this, one just needs to substitute the tree mass 
with the running mass \cite{st97,st97bis}:
\be
V = V_0 + \frac12m^2(\ln \phi)\phi^2 + \dots
\, .
\label{runpot}
\ee
Here $ m^2(\ln \phi) $ is obtained by integrating the RG 
equation of the form
\be
{\diff m^2\over \diff\ln\phi} =
{\diff m^2\over\diff\ln Q} = \beta_m 
\label{rge}
\, ,
\ee
with $\beta_m $ being the $\beta $-function of the soft
inflaton mass and depending on all its couplings.
A crucial assumption of the running mass model is that
the radiative corrections are substantial, as  
they are exploited to realize slow-roll in some region 
of the potential.
In fact at the Planck scale $\phi\sim\mpl$, the 
mass-squared is supposed to have the generic supergravity value 
$|m^2 |\simeq H_I^2 = V_0/(3 \mpl^2)$. The radiative corrections
drive down $m^2$, so that, when $\phi$ is many orders of magnitude below
$\mpl$, it has the much smaller value which is needed for viable inflation.
There are four types of model, depending on the 
sign of $m^2$ at the Planck scale, and on whether or not that sign has
changed by the time that the inflationary regime is reached.

Note that sufficiently strong running is realized for example 
in the case of radiative EW symmetry breaking in the Minimal
Supersymmetric Standard Model, where the running turns one of 
the Higgs doublet's mass from positive to negative: the main 
difference is that in this case it is sufficient to suppress 
the mass and so we do not need to rely on a very large
coupling, as the top Yukawa.
Unfortunately, since some of the fields take a large $\mpl $
v.e.v. after inflation, it is not viable to implement the 
running mass model directly within the MSSM, but possibly 
in some of its extensions~\footnote{For a tree-level example based 
on the GUT group $SU(6)$, see e.g.~\cite{rsg}.}.

In general at one loop $\beta_m $ is given by \cite{st97bis,c98}
\be
\beta_m = - \frac{2 C}\pi \alpha \widetilde m^2
+ {D\over 16\pi^2} |\lambda |^2 m^2_{loop}
\, ,
\label{betam}
\ee
where the first term arises from the gauge interaction with
coupling $\alpha$ and the second from the Yukawa interaction $\lambda $. 
It is easy to generalize to the case of more gauge groups or Yukawas.
In the expression above, $C, D$ are positive group-theoretic numbers 
of order one, counting the degrees of freedom present in the one-loop
diagrams contributing to the running, $\widetilde m$ is the 
gaugino mass, while  $m^2\sub{loop}$ is 
the common susy breaking mass-squared of the scalar particles 
interacting with the inflaton via Yukawa interaction.
Note that the first term in~\eq{betam} is always negative,
while the second has no definite sign, since $m^2\sub{loop}$ 
is defined as the mass squared splitting between scalar and 
fermionic superpartners and can have either sign.
Also the case of a non-interacting inflaton gives directly 
$\beta_m = 0$ and it coincides with the constant mass 
potential. Anyway, in realistic cases the inflaton must have some
interaction in order to reheat the universe or to secure a hybrid 
end to inflation and so in any model we expect naturally some 
running, even if perhaps below the level required by the 
running-mass model~\footnote{For example one could envisage
a very heavy inflaton coupling only gravitationally and in that
case the running would be negligible. Note that we are restricting 
here to models where the $\beta $-function is generated by 
renormalizable interactions.}.

Over a sufficiently small range of $\phi$, or for small inflaton 
couplings, it is a good approximation to take a truncated
Taylor expansion of the running mass $m^2(\ln\phi)$ around 
a particular scale, which we will choose as $\phi\sub{0}$, 
the inflaton value at the epoch of horizon exit for the
pivot scale $k\sub{0} $; for comparison with the WMAP results
\cite{WMAP-peiris} we choose $k\sub{0} = 0.002 h \Mpc^{-1} $.
Then we have:
\be
V=V_0 +\frac12m^2(\ln \phi\sub{0})\phi^2 
-\frac{3}{2} c H_I^2 \phi^2 \;
\ln\({\phi\over \phi\sub{0}}\)
\, ,
\label{vlin1}
\ee
where we have rescaled the last term w.r.t. $3 H_I^2$ 
for future convenience.
The dimensionless constant $c$ is proportional to the 
mass beta function at the particularly chosen point,
\be
c = - {\beta_m (\ln \phi\sub{0}) \over 3 H_I^2} \, .
\label{cofbeta}
\ee
The linear expansion corresponds on the quantum field theory
side to the one--loop expansion, since it practically neglects 
the running of $\beta_m $, which arises at two--loops.
It has been shown~\cite{cl98} that for small $c$, 
as is required by the slow roll conditions, this linear 
approximation is more than sufficient over the range of $\phi$ 
corresponding to horizon exit for astronomically interesting 
scales, i.e. between $k\sub{0}$ and $8h^{-1}\Mpc$. 

To simplify the expressions, it is
very useful to introduce a new parameter $\phi_*$ via
\be
m^2(\ln\phi\sub{0}) = 3 H_I^2 \, c 
\[\ln\({\phi_*\over \phi\sub{0}}\) +\frac12\]
\, .
\label{mofphi}
\ee
Then \eq{vlin} takes the simple form \cite{cl98}
\be
V=V_0-\frac{3}{2} c H_I^2 \phi^2\( \ln\(\frac{\phi}{\phi_*}\)
-\frac12 \) 
\,,
\label{vlin}
\ee
leading to 
\be
\frac{V'}{V_0} = -c\frac{\phi}{\mpl^2} \ln\(\frac{\phi}{\phi_*}\) 
\,,
\ee
In typical cases the linear approximation is valid at $\phi=\phi_*$,
and that point is then a maximum or a minimum of the potential.

The running-mass model supposes that all soft masses at the 
Planck scale (or some other high scale) have magnitude roughly 
of order
\be
\tilde m^2, | m_{loop}^2|  \sim 3 H_I^2
\,,
\label{massmag}
\ee
and that the couplings (gauge or Yukawa) are perturbative,
but not too small for the running to be substantial. 
In general then, in the RG equation, one product of coupling times
mass scale will dominate over the others and be relevant besides 
the inflaton mass. 
Then we expect the coupling  $c$ defined by \eq{cofbeta} to be 
roughly
\be
|c| \sim
\left\{
\begin{array}{c}
\alpha {\tilde m^2 \over 3 H_I^2}\\
\\
|\lambda |^2 { m_{loop}^2 \over 3 H_I^2}
\end{array}
\right\}
\sim 10^{-1}\mbox{ to }10^{-2}
\,.
\label{c-range}
\ee
A bigger value of $|c|$ would not allow slow roll inflation and 
infringe upon the validity of our Taylor expansion \eq{vlin1}. 
On the other hand, using \eq{vlin} as a very crude 
estimate of the inflaton mass at the Planck scale, one can see 
that a much smaller value of $|c|$ is probably not viable
either, since it would require the inflaton mass at that scale 
to be suppressed below the estimate \eq{massmag}. 

The value of $c$ is directly related to the supersymmetry breaking
masses in the scalar and gauge sector and therefore the
theoretical question arises under which conditions we expect
\eq{massmag} to be satisfied.
We will assume that supersymmetry breaking originates from
an F-term not a D-term and that the inflationary scale $V_0$ 
is obtained due to the mismatch between the two contributions 
in the SUGRA potential, the F-term part and the negative part
proportional to the superpotential $ 3 |W|^2/\mpl^2$.
Note that we do not need to assume that the second term vanishes 
(so the gravitino remains all the time massive), but it is
for our purposes sufficient that the F-term during inflation 
is such that
\begin{equation}
H_I \geq m_{3/2}^{infl}.
\end{equation}
The most economical cases are probably if the two quantities
are of the same order and/or if the gravitino mass during inflation
is very near to the vacuum value.

Taking both assumption seriously, we can get some estimate of 
$H_I$, depending on the mechanism for supersymmetry breaking:
\bea
H_I &\sim & 10^{4}\GeV\quad \mbox{anomaly-mediation}\nonumber\\
H_I &\sim & 10^{2}\GeV\quad \mbox{gravity-mediation}\nonumber\\
H_I &\sim & 10^{-3} \GeV\quad \mbox{gauge-mediation} 
\label{H-scale}
\,.
\eea
But note that the assumptions above can be easily relaxed and 
any larger value could be plausible.

In this setting, if the dominant contribution to the scalar
masses comes from supergravity and barring cancellations, we
naturally expect \eq{massmag}.
For what regards the gaugino mass, the expectation of
\eq{massmag} is not so straightforward because gaugino masses 
can be very small with some types of SUSY breaking, depending on 
the gauge kinetic function and its dependence from the moduli fields.
But it is not unreasonable, as long as the inflationary scale
is of the order of the gravitino mass in the vacuum. 

We will investigate in the following how good does the model
compare with the data for reasonable values of $c$ and try to
reach conclusions on the naturalness of the allowed parameters.

\subsection{The spectrum and the  spectral index}

Let us now discuss the more phenomenological issues of the predicted 
spectrum and spectral index of the primordial density perturbation. 
We suppose that it is generated by the inflaton field perturbation,
which means that it  is purely adiabatic and gaussian. It is therefore
specified by the curvature perturbation $\calr(k)$,
with $k$ as usual the comoving wavenumber. This quantity is gaussian and
hence specified by its spectrum $\calpr(k)$.

To express such spectrum in the running mass model, 
it is convenient to define yet another parameter
\bea
s \equiv  c \ln\({\phi_*\over\phi\sub{0}}\) 
\, ,
 \label{sigma}
\eea
where the subscript 0 denotes as before the epoch of horizon exit 
for the pivot scale $k\sub{0} = 0.002 h\Mpc^{-1}$. 

Note that $s $ is directly related to a physical parameter of the
potential by the relation:
\be
s + \frac12 c \equiv {m^2 (\ln\phi_{0}) \over 3 H_I^2} \, ;
\ee
so from $s$ and $c$ we can directly access the inflaton
mass and its couplings at $\phi_{0}$. Note also that the 
inflaton mass at that scale should be smaller than the expected 
value at $\mpl$ and therefore, barring cancellations,
the parameter $s $ is expected to be smaller than unity.
We will later plot our results not only in the $s$ and $c$ plane, 
but also in the physical parameters space $m^2 (\ln\phi_{0})/(3 H_I^2)$
and $- \beta_m(\ln\phi_{0})/(3 H_I^2)$ for fixed value of $H_I$.

At the pivot scale, the 
prediction of the running-mass model is
\begin{eqnarray}
\calp_{\calr}^{1/2} (k\sub{0}) &=& \frac{1}{2 \pi \sqrt{3}} 
\frac{V_0^\frac12}{\mpl |\phi\sub{0}\, s|} \\
 &=& \frac{1}{2 \pi \sqrt{3}} 
\frac{V_0^\frac12}{\mpl |\phi_*\, s|} \exp(s/c)
\label{power1}
\,.
\end{eqnarray}
This quantity is related to the normalization of the CMB
power spectrum by the relation~\cite{WMAP-peiris}
\begin{equation}
\calp_{\calr}^{1/2} (k\sub{0}) = 5.43\times 10^{-5} A^{1/2}
= (4.7\pm 0.5) \times 10^{-5} 
\end{equation}
where we have used $A= 0.75^{+0.08}_{-0.09} $, obtained by the
global WMAP fit~\cite{WMAP-peiris}.
This normalization can be easily satisfied for choices of
$V_0$ and $\phi\sub{0}$ or $\phi_*$ that correspond to 
reasonable particle physics assumptions.
Note that eq.~(\ref{power1}) can also be recast in the form:
\be
|\phi_{0}|\; |s| =  \frac{H_I}{2\pi \calp_{\calr}^{1/2}(k\sub{0})}
\ee
where $s$ also depends logarithmically on $\phi_{0}$ as given
in \eq{sigma}. 
This expression can be also used for estimating $|s|$
\be
|s| =  \frac{1}{2\pi\calp_{\calr}^{1/2}(k\sub{0})} 
\frac{H_I }{|\phi_{0}|}\, ,
\ee
and therefore the parameter $s$ is also directly related to the
value of the inflaton field compared to the inflationary scale.
So for $|s|< 1$ we must have
\be
\frac{H_I }{2\pi\calp_{\calr}^{1/2}(k\sub{0}) } \leq |\phi_{0} | \ll \mpl\, ,
\ee
where the last inequality stems from the requirement of
negligible higher order supergravity corrections.
We see therefore that in this kind of models we expect the
inflationary scale $H_I$ to be much lower than 
$2\pi \calp_{\calr}^{1/2} (k\sub{0}) \mpl \simeq 
1.14 \times 10^{14} $ GeV.
It could be therefore natural to link $V_0 \simeq H_I^2 \mpl^2 $ 
to an intermediate scale like the supersymmetry breaking scale.
Also the inflaton value $ \phi\sub{0} $ must indeed be larger 
than any mass splitting and therefore our use of the RGE-improved
potential is perfectly consistent.

In this paper we would like to constrain the strong 
scale-dependence of the spectrum, which is given by 
\be
\frac{\calp_{\calr}(k)}{\calp_{\calr}(k\sub{0})}
=\exp\[ {2s \over c} \( e^{c \Delta N(k)} -1\)-
2 c \Delta N (k) \]
\,,
\ee
where 
$ \Delta N(k) \equiv N\sub{0} - N(k)
\equiv \ln(k/k\sub{0}) $.

To discuss the spectral index and its running, we need the
first four flatness  parameters \cite{treview,book}, given by
\bea
\epsilon &\equiv& \frac12 \( \frac{\mpl^2 V''}{V} \)^2
\simeq  { s^2 \phi^2\over \mpl^2} e^{c\Delta N(k)}\\
\eta&\equiv& \frac{\mpl^2 V''}{V} \simeq  s  e^{c\Delta N(k)} -c\\
\xi^2 &\equiv& 
\frac{\mpl^4 V' V'''}{V} \simeq -cs e^{c\Delta N(k)} \\ 
\sigma^3 &\equiv & \frac{\mpl^6 V'^2 V''''}{V} \simeq cs^2  e^{2c\Delta N(k)}
\,.
\label{sigma3}
\eea
The parameters are evaluated at the epoch of horizon exit for the scale
$k$. The first parameter $\epsilon$ is negligible because
 $\phi/\mpl$ is taken to be very small. The condition for slow-roll
inflation is therefore just $|\eta|\ll 1$, which is satisfied in the
regime  $\phi \sim \phi_*$ provided that $|c|,|s|\ll 1$, disregarding
the fine-tuned cancellation between the two terms. 

Additional and generally stronger constraints on $s$ follow from the 
reasonable assumptions that the mass continues to run to the end of 
slow-roll inflation, and that the linear approximation remains  
{\em roughly} valid. 
Discounting the possibility that 
 the end of inflation is very fine-tuned, to occur close to the
maximum or minimum of the potential, we have the lower bound 
\be
|s|\gsim e^{-c N\sub{0}}|c|
\,.
\label{sb1}
\ee
Note that for negative $c$, this constraint is very strong,
requiring a very large value of $s$ even for small $c$
and a kind of fine-tuning between $s$ and $c$ to give a 
reasonable value of $n-1$. 

\begin{figure}[t]
\begin{center}
\includegraphics[width=3in,height=3in]{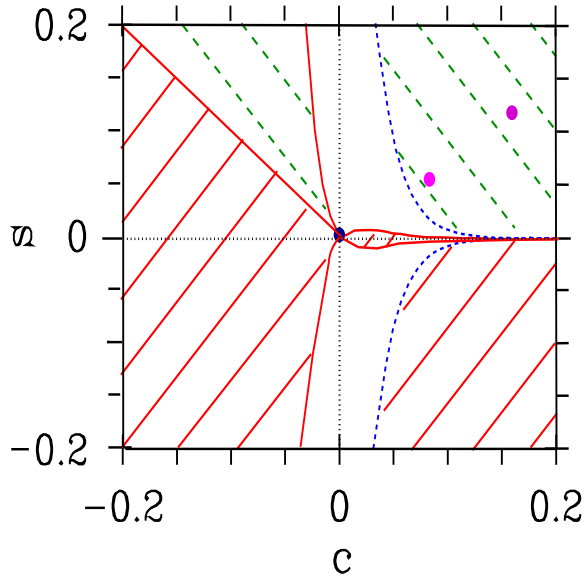}
\caption{The theoretically expected region for the parameters
$s$ and $c$ for a value of $N\sub{0} = 50$; the 
solid(red)-line-hatched region is strongly excluded, while 
the dashed(green)-line-hatched region is excluded only if
the mass is supposed to run up to the end of inflation.
The dotted (blue) line gives the prediction for the case when
the end of inflation is triggered by $\eta = 1$.
The circles correspond to the values in the explicit models
discussed in section II.C: the upper ones (magenta) refer to the
case of gauge coupling dominance, while the (blue) one 
at the origin to the case of Yukawa coupling dominance.}
\label{theory-plot}
\end{center}
\end{figure}

For positive $c$, we also obtain a significant upper bound
by setting $\Delta N= N\sub{0}$ in \eq{runpred}, and remembering
that slow-roll requires $|n-1|\lsim 1$:
\be
|s|\lsim e^{-c N\sub{0}}\hspace{2em}(c>0)
\,.
\label{sb2}
\ee
In the simplest case, if slow-roll inflation 
ends when $n-1$ actually becomes of order 1, this bound becomes
an actual estimate, $|s|\sim e^{-cN\sub{0}}$.
As discussed in \cite{cl98}, this upper bound can be relaxed for
positive $s$ if the running of the mass ceases before the end
of slow-roll inflation.
The approximate region of the $s$ versus $c$ plane excluded by these
considerations is shown in Figure~\ref{theory-plot}.

Since $\epsilon$ is negligible, the spectral index to second
order is
\bea
n(k) &=& 1 + 2\eta \left( 1+ {\eta\over 3} \right) + 2.13\,\xi^2 
\,.
\eea
This gives \cite{cl98}
\begin{eqnarray}
{n(k)-1\over 2} &=& 
s e^{c\Delta N(k)} \left( 1- 1.06 c\right) - c \nonumber\\ 
& & + {1 \over 3}\left( s e^{c\Delta N(k)} - c \right)^2
\,.
\label{runpred}
\end{eqnarray}
The first derivative of the spectral index is given by
\be
n'(k) \equiv \frac{\diff n (k)}{\diff \ln k} 
= 2 s c\, e^{c \Delta N(k)}
\,.
\label{diffn}
\ee
Clearly the spectral index is not constant within cosmological 
scales unless $s$ or $c$ is very close to zero.
Note that for this type of models, $n'$ is a higher order effect
(suppressed by both $c$ and $s$) as in usual slow-roll inflation, 
but it is not proportional to $(n-1)^2$. So it is perfectly 
allowed to have $n=1$ at a particular scale, with a non-zero running.
This is again a consequence of our initial assumption that in
the region of the potential where inflation takes place the
one-loop contribution to the inflaton mass is of the same order 
as the tree one. Note anyway that the perturbative expansion is 
not endangered by our assumption, since the higher orders stay
always smaller than the one-loop. 
Similarly for the slow-roll expansion, the second order can become
larger than the first one in case of strong cancellation between
 $s$ and $ c$, but the perturbative expansion is still 
solid.

At the 
pivot scale, we have, for example, to second order in $s,c$  
\bea
n\sub{0}-1 &=& 2 (s-c-1.06 s c) + {2\over 3} (s-c)^2\\
n'\sub{0} &=&  2sc
\label{n-dn}
\,.
\label{ncobe}
\eea
The line $s = c/(1-1.06 c)$ in the $s$ vs $c$ plane
corresponds approximately to $n\sub{0}= 1$. 
The Harrison-Zeldovich case of constant $n=1$ is given by the 
origin $s=c=0$, while constant spectral index different from~$1$ 
is realized either near the $c=0$ axis for $s = (n-1)/2$ or 
near the $s=0$ axis for $c = - (n-1)/2$. 

Since the phenomenological parameters only depend on $s-c$
and $sc $, as long as higher orders are negligible,
the allowed region is expected to be symmetric under reflection
along the $s+c=0$ line.
We can solve the system of equations exactly and extract the 
parameters $s$ and $c$ from a measurement of $n\sub{0}$ and 
$ n'\sub{0} $; one solution is given as
\bea
c_1 &=& - {1\over 4 } \left[ n\sub{0}-1 + 1.06 n'\sub{0}
\right.\\
& & \left. - \sqrt{\left(n\sub{0}-1 + 1.06 n'\sub{0} \right)^2 
+ 8 n'\sub{0}}\right] .\nonumber\\
s_1 &=&   {1\over 4 } \left[ n\sub{0}-1 + 1.06 n'\sub{0}
\right.\\
& &  \left.+ \sqrt{\left(n\sub{0}-1 + 1.06 n'\sub{0} \right)^2 
+ 8 n'\sub{0}}\right] \nonumber\; .
\eea
the second solution is given just by the symmetry, i.e.
$ c_2= -s_1 $ and $s_2 = -c_1$.

From the expressions above, it is clear that not all values of 
$n'\sub{0} $ are allowed in the running mass model:
we obtain the constraint
\be
n'\sub{0} \geq - {(n\sub{0}-1)^2\over 4}
\, ,
\label{bound-dndlnk}
\ee
so that a decreasing spectral index is possible only if $n\sub{0}$
is different from 1. So in general the prediction of the running mass
model tends toward positive $n' $, contrary to the result claimed by 
WMAP~\cite{WMAP-peiris}.
Both due to this constraint and the exponential dependence on $c$,
we see that fitting for arbitrary value of $n'$ is not equivalent 
to performing a fit for the running mass model.
Note also that, as discussed in \cite{clm03}, the fact that 
the running mass model tends in general to give more power 
at low scales for sufficiently large $c$, can give naturally
a large value of the reionization redshift in the Press-Schechter 
approximation and accommodate easily the value obtained by
WMAP \cite{spergel}.

We will show in the following the allowed region both in the
$s, c$ parameter space and in the $n'\sub{0} $ 
vs $n\sub{0}-1$ plane. 

\subsection{Explicit minimal models}

1. Dominance of the gauge interaction

The first model to be proposed \cite{st97,st97bis} considered
the case of an inflaton charged under an asymptotically
free $SU(N)$ group. We will consider the case of two matter 
superfields in the adjoint representation as in \cite{c98}, 
where one has very simply a superpotential given by
\be
W = g S\, \Tr \left(\phi_1\phi_2 \right)\; ,
\ee
then e.g. the direction $\phi = \phi_1^1$ is D- and F-flat 
for vanishing other fields. Note that here for the case of
universal SUSY breaking masses, the role of the waterfall
field can be played by the singlet $S$~\cite{c98} and the
gauge symmetry is unbroken in the true vacuum.

We can then write easily the $\beta $-functions,
\be
\beta_m = - {2 N \over \pi} \alpha \tilde m^2 \; 
\quad\quad \tilde m \propto \alpha\, , 
\ee
while for the gauge coupling
\be
{d \alpha \over d \ln Q} = - {N \over 2 \pi} \alpha^2 \, , 
\ee
giving
\be
\alpha (\ln\phi) = {\alpha (\mpl) \over 1 + {N \alpha_0 \over 2\pi} 
\ln\phi}.
\ee
So the RG equation for the soft mass can be solved
analytically to give
\begin{eqnarray}
m_{\phi}^2 \(\ln \phi\) &=&  m_{\phi}^2 (\mpl) - 2 \tilde m^2 (\mpl)
\nonumber\\ 
& &  
+ {2 \tilde m^2 (\mpl)\over (1+{N\over 2\pi} \alpha (\mpl) \ln\phi)^2} \; ,
\end{eqnarray}
where we have defined the boundary conditions at 
$\mpl $~\footnote{Any function evaluated at $\mpl $ is simply the
initial value and we use $\ln \phi $ to mean $\ln (\phi/\mpl )$.}.

It is then clear that the inflaton mass increases for decreasing 
$\phi < \mpl $ and therefore to obtain a reduction of the mass 
absolute value a negative initial mass is necessary.

For the linear approximation, we have then
\bea
c &=& {2 N \over \pi} \alpha \(\ln \phi_{0}\) \; 
    {\tilde m^2 (\ln \phi_{0}) \over 3 H_I^2} \\
  &=& {2 N \alpha (\mpl) \over \pi} 
{\tilde m^2 (\mpl) \over 3 H_I^2}
{\alpha^3 (\ln \phi_{0}) \over \alpha^3(\mpl)} \, ; 
\eea
which is a positive number, and
\bea
s + {1\over 2} c &=&  {m_\phi^2 (\ln\phi_{0}) \over 3 H_I^2}\\
                 &=&  {m_\phi^2 (\mpl) - 2 \tilde m^2(\mpl) \over 3 H_I^2} 
\nonumber\\
                 & & + {2 \tilde m^2(\mpl) \over 3 H_I^2} \;
                       {\alpha^2 (\ln \phi_{0}) \over \alpha^2(\mpl)}
\eea
so $s$ can be positive or negative.

The power spectrum normalization gives us instead:
\be
\ln\(\phi_{0}\) + \ln (2 \pi |s|) = \ln\(\calp_{\calr}^{1/2} (k\sub{0}) H_I\).
\label{phicobe}
\ee
The expression above can provide directly an estimate of the 
order of magnitude of $\phi_{0} $, but unfortunately it is
not possible to solve directly for this quantity.
We can instead turn the formula around and use it to define the
inflationary scale, after we have singled out the region where
the spectral index is small. 

For $\phi_{0} $ we can use as a very rough estimate instead 
the scale where $m^2_\phi $ exactly vanishes, $\phi_{m=0} $: 
\be
\ln\(\phi_{m=0}\) \simeq {-2\pi \over N \alpha (\mpl)} 
\left[ 1- \left(1+{|m^2_{\phi} (\mpl)| \over 
2 \tilde m^2 (\mpl)} \right)^{-1/2} \right];
\ee
depending on the value of the coupling constant, the scale
$\phi_{m=0} $ changes very strongly. Note that in order for
$n\sub{0} $ to be phenomenologically acceptable, $\phi_{0}$ 
must be not much far away: the value of the spectral index
at $m^2_\phi=0$ is in fact already small,
\be
n_{m=0}-1 \simeq - 3 c\; .
\ee

Assuming $\alpha (\mpl) \simeq 1/24 $ as in SUSY-GUT models, $N=3$ and 
$|m_\phi^2 (\mpl)| = \tilde m^2 (\mpl) = 3 H_I^2 $, we obtain for example
\be
\phi_{m=0} = 9.9 \times 10^{-5} \mpl\, ;
\ee
so that the inflationary scale must therefore be
\be
H_I \sim 10^{-9} \mpl \sim 10^{9} \GeV.
\ee
For the parameters $c, s$ at that point we obtain the values:
\bea
c_{m=0} &\simeq &  0.15 \\  
s_{m=0} & \simeq & -0.07     
\eea
as expected. This corresponds to 
\be
n(k_{m=0})-1 = -0.44
\ee
too large to be phenomenologically acceptable, but e.g for
$\phi_{0} = 0.25\,\phi_{m=0} $ one has 
\bea
c &\simeq &  0.16 \\  
s & \simeq & 0.13     
\eea
and therefore
\begin{eqnarray}
n(k_{0})-1 &=& -0.06\\
n'(k_{0}) &=& 0.04\; .
\end{eqnarray}
We see clearly that the spectral index can be very small, but with
a running of the same order. 

Note anyway that changing $\alpha (\mpl) $ by only a factor 
$1/2$, we obtain much smaller values:
\be
\phi_{m=0} = 9.74 \times 10^{-9} \mpl\, ;
\ee
so that the inflationary scale must therefore be
\be
H_I \sim 10^{-13} \mpl \sim 10^5 \GeV.
\ee
This strong dependence on the coupling is characteristic of
dimensional transmutation and allows us to construct viable
models spanning a very large range of $H_I$, while presenting
similar values for the spectral index and its running.
It is therefore possible to accommodate practically all
the scales in eq.~\ref{H-scale}, and in general smaller
$H_I$ corresponds to smaller gauge coupling, which for
the same gaugino mass means smaller $c$.
In fact in this case we have at $\phi_0 =\; 0.25 \phi_{m=0} $:
\bea
c &\simeq &  0.08 \\  
s & \simeq & 0.06     
\eea
and therefore
\begin{eqnarray}
n(k_{0})-1 &=& -0.02\\
n'(k_{0}) &=& 0.01\; .
\end{eqnarray}

\smallskip

2. Dominance of the Yukawa coupling

We consider here the simplest of the cases studied in \cite{c98}, 
where the superpotential is given by
\be
W = \lambda \phi_1 \phi_2 \phi_3
\ee
and all fields are singlet under gauge interaction.
If the trilinear susy breaking coupling vanishes, we have:
\bea
{d \lambda \over d\ln Q} &=& {3 \lambda \over 16 \pi^2} |\lambda |^2\\
{d m_i^2 \over d\ln Q} &=& {|\lambda |^2 \over 8 \pi^2} \sum_j m_j^2 \; .
\eea
Then for the average scalar mass $\bar m^2 = \sum_i m_i^2/3 $, 
we obtain the simple expression:
\be
\bar m^2 (\ln \phi/\mpl) = {\bar m^2 (\mpl) \over 
             1-{3\over 8\pi^2 } \lambda^2 (\mpl) \ln \phi} \; . 
\ee
The mass differences instead are constant, so for the single masses
we have:
\be
m_i^2 (\ln \phi) = \bar m^2(\ln \phi) 
                         - \bar m^2 (\mpl)+ m^2_i (\mpl) \; .
\ee
Note that in this case the masses run from positive to negative
and that the running is strong at the beginning and then tends
to flatten out at the value given by $\bar m^2 = 0 $. In fact
in this case the Yukawa is non-asymptotically free and tends
to zero at small scales. So in the plateau region, if 
$ m^2 (\mpl) \simeq \bar m^2 (\mpl) $, we have automatically
a flat potential.

In this specific model, slow roll inflation can be realized in 
any of the 3 field directions, but we have to consider some 
non-universal initial masses, since the hybrid end is assured only
if one of them becomes negative, but not all at the same time. 
For example assume $\phi_1$ to be the inflaton and some non-universal 
mass terms coming from higher order term in the K\"ahler potential so
that $m_1^2(\mpl) = H_I^2$, but $ m_2^2(\mpl) + m_3^2(\mpl) = 5 H_I^2$.
Then we have for the inflaton mass
\be
m^2 (\ln \phi ) = {2 H_I^2 \over 
             1-{3\over 8\pi^2 } \lambda^2 (\mpl) \ln \phi} - H_I^2
\ee
and in the linear approximation this gives
\bea
c &=& - \frac{\lambda^2(\phi_{0})\,\bar m^2(\phi_{0})}{8\pi^2\, H_I^2}
\nonumber\\
  &=& - \frac{\lambda^2 (\mpl)}{12\pi^2} { 1 \over 
             \(1-{3\over 8\pi^2 } \lambda^2 (\mpl) \ln \phi_{0}\)^2} \\
s &=& - \frac{c}{2} + \frac{2}{3} \[ {1 \over 
             1-{3\over 8\pi^2 } \lambda^2 (\mpl) \ln \phi_{0}} 
             - \frac{1}{2} \] \; ;\nonumber\\
\eea
so that $c$ is negative, while $s$ can have either sign.
We note that in this type of models, the parameter $c$ is related 
to second power of the coupling constant and therefore $\lambda $
has to be sufficiently large to give an effect.

As long as $s$ does not vanish in the interesting region, 
we can find an estimate of $\phi_{0} $ from the COBE 
normalization. Since the running is slower in this case, we
can solve iteratively eq.~(\ref{phicobe}) for $\phi\sub{0} $ 
and $s$, taking $H_I = 10^{-15} \mpl \sim 10^3 \GeV $ 
and $\lambda(\mpl)  =1 $.
We obtain then
\be
{\phi\sub{0}\over \mpl} =  1.3 \times  10^{-10}
\ee
and
\bea
c &=& -0.002 \\
s &=& 0.026\; .
\eea
We can in this case have $|c| < 0.01$ since we have assumed
a somewhat suppressed initial inflaton mass 
$m^2(\mpl) = H_I^2 < 3 H_I^2 $.
So in this case we have very small scale dependence:
\bea
n(k\sub{0})-1 &=& 0.056  \\
n'(k\sub{0}) &=& - 0.0001.
\eea

\medskip
\section{Observational constraints}
\medskip

\subsection{Method}

Our analysis method is based on the computation of a likelihood 
distribution over a grid of precomputed theoretical models. 

We restrict our analysis to a flat, adiabatic, $\Lambda$-CDM model 
template computed with a modified version of CMBFAST \cite{CMBFAST}, 
sampling the parameters as follows: 
$\Omega_{cdm}h^2\equiv \omega_{cdm}= 0.05,...0.20$, in steps of $0.01$; 
$\Omega_{b}h^2\equiv\omega_{b} = 0.0018, ...,0.030$, in steps of $0.001$ and 
$h=0.55, ..., 0.85$, in steps of $0.05$. 
The value of the cosmological constant $\Lambda$ is determined by the 
flatness condition.
Our choice of the above parameters is motivated by Big Bang Nucleosynthesis 
bounds on $\omega_b$ (both from $D$ \cite{burles} and $^4He+^7Li$ 
\cite{cyburt}), from supernovae \cite{super1} and galaxy clustering 
observations (see e.g. \cite{thx}). 
Current data also does not show evidence for additional physics
like quintessence (see e.g. \cite{bean}), extra-relativistic
particles (see e.g. \cite{bowen}), topological defects
(see e.g. \cite{dkm}) or isocurvature perturbations (see e.g. \cite{jxd}).
We do not consider massive neutrino which may have an effect on
our results but that are probably negligible (see e.g. \cite{fogli}).
From the grid above we only consider models with age of the universe 
$t_0>11$ Gyrs.
We allow for a possible (instantaneous) 
reionization of the intergalactic medium by varying the reionization redshift 
$5 < z_{ri}< 25$  and we allow a free re-scaling of the fluctuation 
amplitude by a pre-factor of the order of $C_{110}$, in units of 
$C_{110}^{WMAP}$ as measured by the WMAP satellite. 
Finally, we let the running parameters $c$ and $s$ vary as follows: 
$-0.2 < c < 0.2$, and $-0.2 < s < 0.2$ in steps of $0.008$.

The theoretical models are compared with the recent temperature and 
temperature-polarization WMAP data using the publicly available likelihood 
code \cite{WMAPanalysis03}.

In addition to the CMB data we also consider the constraints on the 
real-space power spectrum of galaxies from the SLOAN galaxy
redshift surveys using the data and window functions of the analysis
of ~\cite{Tg04}. We restrict the analysis to a
range of scales over which the fluctuations are assumed to be in the
linear regime ($k < 0.2 h^{-1}\rm Mpc$). When combining with the CMB
data, we marginalize over a bias $b$ considered as 
an additional free parameter.

We also include information from the Lyman-alpha Forest in the 
Sloan Digital Sky Survey using the results of the analysis of
\cite{Se04} and \cite{Mc04} which probes the amplitude
of linear fluctuations at very small scales. 
For this dataset, small-scale power spectra are computed 
at high redshifts and compared with the values presented in 
\cite{Mc04}.

\medskip
\subsection{Results}
\medskip

\begin{figure}[h!]
\begin{center}
\includegraphics[width=2.7in]{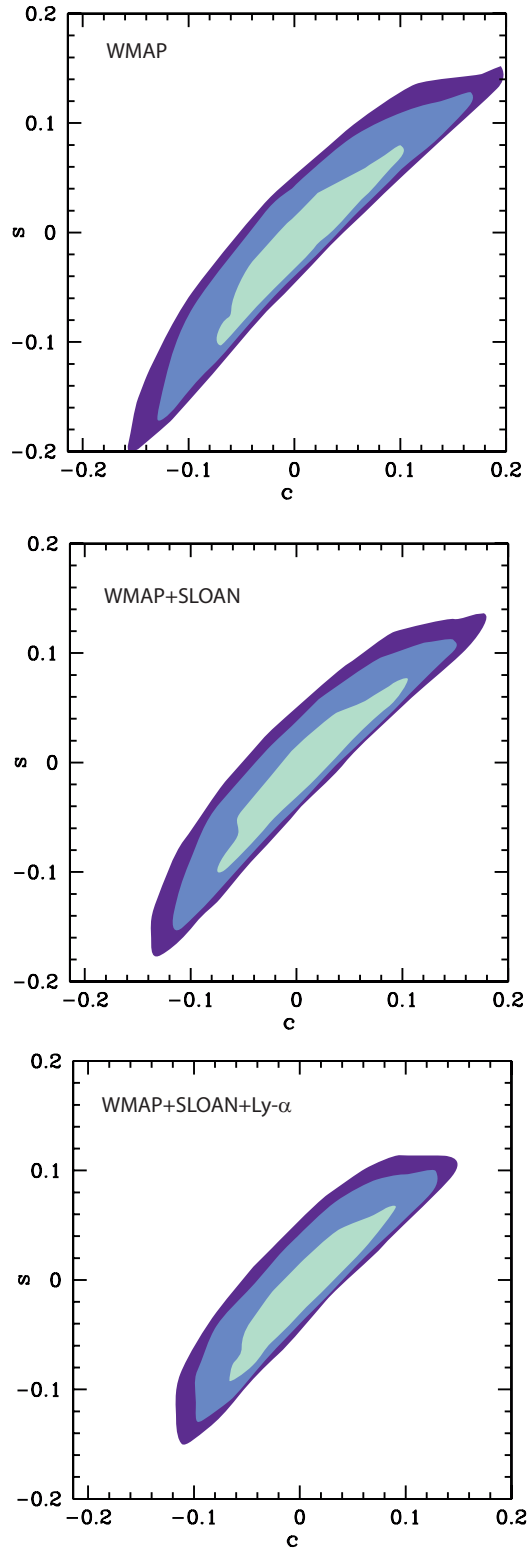}
\caption{Likelihood contour plot in the $c-s$ plane showing the 
$1$,$2$ and $3\sigma$ contours from the WMAP data (Top Panel),
WMAP+SLOAN (Middle Panel) and WMAP+SLOAN+Ly-$\alpha$ (Bottom Panel).}
\label{figlike}
\end{center}
\end{figure}

\begin{figure}[t]
\begin{center}
\includegraphics[width=5.5in]{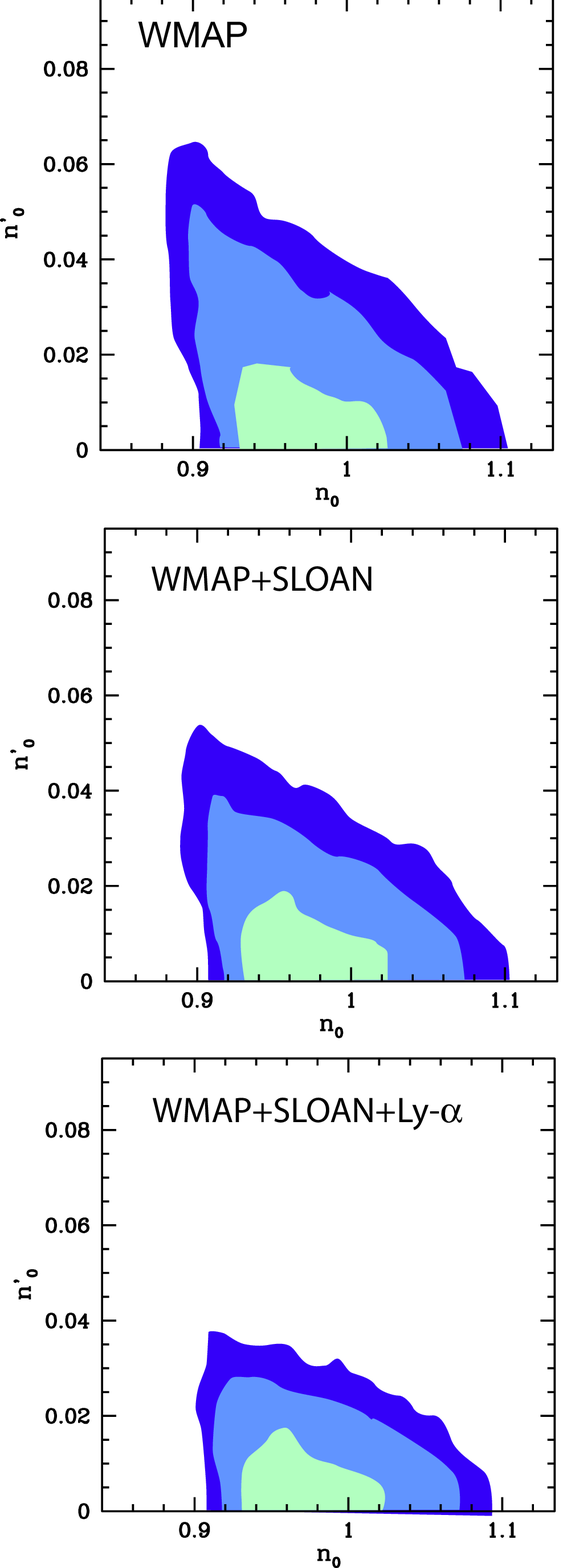}
\caption{Likelihood contour plot in the $n\sub{0}-n'\sub{0}$ 
plane showing the $1$,$2$ and $3\sigma$ contours from the WMAP data
(Top Panel),
WMAP+SLOAN (Middle Panel) and WMAP+SLOAN+Ly-$\alpha$ (Bottom Panel).}
\label{figlike2}
\end{center}
\end{figure}

\begin{figure}[t]
\begin{center}
\includegraphics[width=3in]{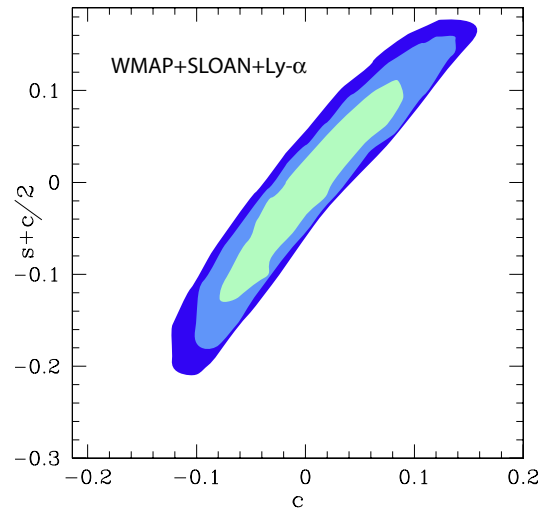}
\caption{Likelihood contour plot in the plane 
$c-(s+c/2)$  showing the $1$,$2$ and $3\sigma$ contours from the 
WMAP+SLOAN+Ly-$\alpha$ data. We recall that these parameters
are related to the physical inflaton potential parameters
by $ c = -\beta_m(\ln\phi_{0})/(3 H_I^2) $ and 
$ s+c/2 = m^2(\ln\phi_{0})/(3 H_I^2) $.
}
\label{figlike3}
\end{center}
\end{figure}

In Fig.~\ref{figlike} we plot the likelihood contours in the 
$c-s$ plane showing the $1$, $2$ and $3\sigma$ contours. 
The top panel is WMAP, middle WMAP+SLOAN and bottom panel
WMAP+SLOAN+Ly-$\alpha$.
As we can see there is a strong correlation between the two 
parameters along the $c-s$ direction and the inclusion of
the SLOAN data does not improve significantly the CMB constraints.
However adding the Lyman-$\alpha$ datasets breaks the degeneracy and 
shrinks the likelihoods. As already noticed by \cite{Se04}, we find 
that the Lyman $\alpha$ data are able to restrict more strongly the scale 
dependence of the spectral index and therefore exclude the
parameter space at large $|c|$; in particular at $68 \%$ we have
$ sc < 0.0043 $.

Focusing on the region ($s+c>0$),
we plot the likelihood contours in the $2(s-c)+1$ vs. $2sc$ plane in 
Figure~\ref{figlike2}. As we explained before, $2(s-c)+1$ gives 
the value of the spectral index $n_0$, while $2sc= n'_0 $ gives 
the bend in the spectrum. Note that due to the bound (\ref{bound-dndlnk}),
the viable negative running region is practically indistinguishable
from the $n_0'=0 $ axis in our scale and therefore we do not show it. 
The maximal negative running is in fact $-0.0015 $ at $95 \%$ c.l. and
it does not basically change with the different datasets. 
As we can see from the figure, the WMAP data alone constrains
$n'\sub{0} < 0.05$ at $95 \%$ C.L.. Including the constraints from 
SLOAN and Ly-$\alpha$ limits the amount of deviation from scale 
invariance to $n'\sub{0} <0.024$ at $95 \%$ c.l..

Finally, in Fig.\ref{figlike3} we plot the likelihood contours in the 
$c$ vs. $s+c/2$ plane. As discussed previously, this variables
correspond to the physical parameters in the inflaton potential
rescaled by the inflationary scale $3 H_I^2$.
It is clear from the graph that the data require a correlation
between the inflaton mass and the $\beta$-function for large
$c$ in order to give a small $n_0 -1 $. It is questionable if
such correlation corresponds to a fine-tuning, and in general
depends on the explicit realization of the model.
In fact in the gauge dominated case, we have already emphasized 
that there is relatively large freedom, since the physical 
parameters are more than the observables. For the Yukawa dominated
case the situation is more constraint, also because the running
is weaker and to be effective requires always large couplings.
In Figure~\ref{theory-plot} are shown  the three points in the 
parameter space that we looked at in detail in Section III C. 
We see that WMAP+SLOAN datasets are not able to exclude any of the models,
but the inclusion of Ly-$\alpha$ excludes the first model discussed 
at $99 \%$ c.l..

\medskip
\section{Conclusions}
\medskip

The rather full analysis that we have described confirms the general
picture indicated by previous analysis \cite{covi,clm03}.
The allowed region in the $c$ vs $s$ plane depicted in 
Figure~\ref{figlike} should be compared with the region  shown in
Figure~\ref{theory-plot} which 
approximately delineates the theoretically disfavoured  region,
and also with the minimum value $|c|\sim 10^{-2}$ which is probably 
needed to generate enough running of the mass
even if we go from the Planck scale to $100\GeV$. 
Combining all of these,
we see that if $|c|$ is significantly above the minimum value,
only the version of the model with $c$ and $s$ both positive
is viable. In that case, the spectral index has significant running which
will be detectable in the foreseeable future. On the other hand, 
if $|c|$ is really of order $10^{-2}$,
all choices of the signs of $c$ and $s$ are possible except maybe
negative $c$ with positive $s$. Furthermore, if that extreme case can be
realized in a viable running-mass model
the running of $n$  will be so small that it may never be detectable.
The data are now starting to squeeze the allowed region to values 
$|c| \leq 0.1 $, still away from the lower bound $10^{-2}$.
The smallness of $|c|$ could be interpreted as a hint that the 
inflationary scale needs probably to be low, to make the running 
from $\mpl$ effective. Note anyway that $H_I$ values of the order of 
$100\GeV$, the expected soft SUSY breaking masses 
in the true vacuum from gravity mediation, are still acceptable,
as demonstrated in the cases of the simple models presented.

Looking at the observational situation in more detail, our
 results show again that the CMB data can put very strong
constraints on the value of the spectral index at large scales,
$n_{0} =1 + 2(s-c)$, but still allow a pretty large scale-dependence.
Other information on the power spectrum, like Lyman $\alpha$ data,
are needed to reduce the parameter space in the orthogonal 
direction.
Even with this inclusion, though, values of $|c|$ of the order of 0.1 
are allowed and we have $n'\sub{0} \leq 0.037 $ at $99 \%$ c.l..
Our allowed region also looks still symmetric under 
reflection with respect to the $s+c=0$ line: this means that the present 
data are not sensitive enough to distinguish the variation of $n'$
that is predicted by the running-mass model.

\textit{Acknowledgments} 

LC would like to thank D. Schwarz and C. Scrucca for
useful discussions.
CJO is supported by a Marie Curie Intra-European Fellowship, grant $501007$.

\end{document}